\documentclass[11pt]{article}
\usepackage{latexsym}
\usepackage{graphicx}
\usepackage{psfrag}

\newcommand{\sect}[1]{\setcounter{equation}{0}\section{#1}}
\newcommand{\subsect}[1]{\subsection{#1}}
\newcommand{\subsubsect}[1]{\subsubsection{#1}}

\newfont{\extra}{msbm10 scaled\magstep1}

\newcommand{\extr}[1]{\mbox{\extra #1}}

\def\d(#1){\partial_{#1}}

\def\R{\extr R}

\def\Z{\extr Z}

\def\be{\begin{equation}}
\def\ee{\end{equation}}
\def\bea{\begin{eqnarray}}
\def\eea{\end{eqnarray}}

\def\p{\partial}
\def\a{\alpha}
\def\b{\beta}

\def\g{\gamma}

\def\l{\lambda}

\def\om{\omega}

\begin{document}

\begin{center}{ \LARGE \bf
Superintegrable  quantum u(3)--systems and\\[2ex]
higher rank factorizations}
\end{center}
\vskip0.25cm

\begin{center}
J.A. Calzada$^{\dagger}$, J. Negro$^*$, M.A. del Olmo$^*$
\vskip0.15cm

${}^{\dagger}${\it Departamento de Matem\'atica Aplicada\\
$^{*}$Departamento de  F\'{\i}sica Te\'orica, At\'omica y \'Optica\\
 Universidad de  Valladolid,
 E-47005 Valladolid,  Spain}
\vskip0.15cm

E. mail:
{juacal@eis.uva.es},
{jnegro@fta.uva.es},
{olmo@fta.uva.es}

\end{center}

\vskip0.5cm
\centerline{\today}
\vskip0.5cm

\begin{abstract}
A class of two-dimensional superintegrable systems on a constant curvature surface is considered as the natural generalization of some well known one-dimensional factorized systems. By using standard methods to find the shape-invariant intertwining operators we arrive at a $so(6)$ dynamical algebra and its Hamiltonian hierarchies. We pay attention to those associated to certain unitary irreducible representations that can be displayed by means of three-dimensional polyhedral lattices. We also discuss the role of superpotentials in this new context.
\end{abstract}


\sect{Introduction}

This work deals with a class of superintegrable Hamiltonians systems, in the framework of
the Schr\"odinger equation of quantum mechanics, and its connections with the
factorization method. We will restrict ourselves to a particular case
where the underlying symmetry is the Lie algebra $u(3)$, but its main features
can be directly implemented to higher dimensional systems.

The main objective of this study is to show a natural extension  to higher
dimensional spaces of the intertwining (or Darboux) transformations
 from a well known class of one-dimen\-sio\-nal factorized systems. In fact, we
want to set the higher rank $u(n)$-systems corresponding to those
having as  dynamical algebra the Lie algebra  of rank one $u(2)$. We will show in detail that the application of procedures familiar in one dimension to a concrete
two-dimensional system will lead us to a wide set of operators closing a dynamical Lie algebra. We  also consider discrete symmetry operators
quite important to perform equivalences. All these operators connect
eigenstates that can be drawn as  points in a three-dimensional lattice
giving rise to polyhedrons representing degenerate series of $u(3)$ irreducible representations. 
Each of these series corresponds  to the same energy  and can be embedded in just one representation of the Lie algebra $so(6)$.

The notion of superpotential will also be re-examined inside the higher rank
formalism. Thus, the usual procedure to look for solutions with separable
variables can be better appreciated under this point of view.

Thus, we try to implement  the program of generalization
of the factorizable one-dimensional systems involving  Lie
algebras of rank one as dynamical algebras, as can be seen, for instance, in the classical paper by Infeld and
Hull \cite{infeld}. We  also hope that this work will be useful when dealing with other integrable systems, but not necessarily maximally integrable, for instance not enjoing for
such a wealth of factorizations, or even not having a  system of separable variables, but still allowing for algebraic methods \cite{kuru1,kuru2,samani,ranada,santander,ioffe}.

The organization of the paper is as follows. In section~\ref{su3hamiltoniansystem} we will
introduce a two-dimensional superintegrable system and find some separable solutions by standard procedures. Although these polynomial solutions are known and can be found in other references, this will serve us to recall some aspects of the usual factorization technique and to precise the operators of the  Lie algebra $u(2)$ related with the  Lie algebra $so(4)$, and their support spaces. Next, in  section~\ref{dynamicalsymmetries}, we will look for other sets of intertwining operators, corresponding to the Lie algebras $u(3)$ and $so(6)$, taking also into account discrete symmetries. We characterize the eigenfunctions belonging to irreducible
representations that will be depicted as the points on octahedrons and the interpretation of some
of its planar sections. The analog of the
superpotentials and their relation with certain types of solutions will be considered in 
section~\ref{egeinstatesandfactorizations}. Some conclusions and perspective for future work will close the paper.


\sect{A superintegrable $u(3)$--Hamiltonian system}\label{su3hamiltoniansystem}

We will fix our attention on a superintegrable Hamiltonian system  defined inside a three dimensional Euclidean ambient space \cite{evans,winternitz,calzada,pogosyan}. In
fact, our system  lives on the 2-sphere $\cal S$
\[\label{esf}
{\cal S}\equiv (s_0)^2+  (s_1)^2 +  (s_2)^2 = 1,\qquad
(s_0,s_1,s_2)\in \R^3
\]
In the frame of the Schr\"odinger equation, this Hamiltonian takes the form
\be\label{hh}
        H =  - \left(J_0^2 + J_1^2 + J_2^2 \right)
      +\frac{l_0^2-1/4}{(s_0)^2} + \frac{l_1^2-1/4}{(s_1)^2}+
        \frac{l_2^2-1/4}{(s_2)^2}
\ee
where $(l_0,l_1,l_2)\in \R^3$, and $J_i =-\epsilon_{ijk}s_j\p_k$ (note that the $J_i$'s operators generate the rotation Lie algebra $so(3)$).
We can parametrize $\cal S$ by means of spherical
coordinates $(\phi_1,\phi_2)$ around the $s_2$ axis  given by
\be\label{coo}
   s_0 = \cos \phi_2\,\cos\phi_1,\qquad
   s_1 = \cos\phi_2\, \sin\phi_1,\qquad
   s_2 \,= \sin\phi_2
\ee
Then, the eigenvalue problem
\[
H\,\Phi = E\, \Phi
\]
after substituting the coordinates (\ref{coo}), takes the form of a
separable differential equation
\be\label{h}
\left[ -\d (\phi_2)^2 +
    \tan\phi_2 \d(\phi_2) +
    \frac{l_2^2{-}1/4}{\sin^{2}\phi_2}
 + \frac{1}{\cos^{2}\phi_2}
\left[ - \d(\phi_{1})^2 +
    \frac{l_0^2{-}1/4}{\cos^{2}\phi_1} +
    \frac{l_1^2{-}1/4}{\sin^{2}\phi_1}\right]
\right]\Phi = E\, \Phi
\ee
The solutions separated in the
variables $\phi_1$ and $\phi_2$, i.e.
\[
{\bf  \Phi}(\phi_1,\phi_2) =f(\phi_1)g(\phi_2)
\]
 after replacing in (\ref{h}) originate the equations
\bea
&&\left[ - \d(\phi_{1})^2 +
    \frac{l_0^2-1/4}{\cos^{2}\phi_1} +
    \frac{l_1^2-1/4}{\sin^{2}\phi_1}\right] f(\phi_1) =
\alpha\,f(\phi_1)
    \label{sisn21}\\[0.3cm]
&& \left[ -\d(\phi_2)^2 +
    \tan\phi_2\d(\phi_2) + \frac{\alpha}{\cos^{2}\phi_2}+
    \frac{l_2^2-1/4}{\sin^{2}\phi_2} \right]g(\phi_2) =
    E\,g(\phi_2) \,\label{sisn22}
\eea
where $\a$ is a separating constant. Next we will solve each of these two
equations through standard factorizations giving rise to polynomials. The key
point is that the results obtained for the first equation will match in a certain way
with those of the second one originating degenerate levels.


\subsect{The $\phi_1$--factorization} \label{thephi1--factorization} 

The one-dimensional Hamiltonian (\ref{sisn21}) in the variable $\phi_1$ is a well known example in the theory of factorizations \cite{infeld}. So, in the following we will restrict ourselves to give a list of the
relevant results. We will see later, in section \ref{dynamicalsymmetries}, how to make use of these
considerations in a broader context.

The second order operator at the l.h.s.\ of eq.\ (\ref{sisn21}) can be cast  as a
product of first order operators
\[
H_{(0)}^{\phi_1} = A^{+}_{0}\,A^{-}_{0} + \lambda_0
\]
being
$$
A_{0}^{\pm} =\,\pm \d(\phi_1) -  (l_0+1/2)\,\tan \phi_1 +
(l_1+1/2)\,\cot \phi_1  \qquad
\lambda_0 \,= (l_0 + l_1+1)^2
$$
These elements are part of a family of operators
$\{A_{m}^{+},A_{m}^{-}\,,\lambda_{m}\,,H_{(m)}^{\phi_1}\}$, $m\in
\Z$, where
\bea
&&A_{m}^{\pm} = \pm \d(\phi_1) -  (l_0 + m+1/2)\,\tan \phi_1 +
       (l_1+m+1/2)\,\cot \phi_1  \label{aa}\\[0.2cm]
&&\lambda_{m} =  (l_0 + l_1 + 2\,m+1)^2\,\nonumber \\[0.2cm]
&& H_{(m)}^{\phi_1}=  -\d(\phi_1)^2
    +\frac{ (l_{0}+m)^2 -1/4}{{\cos}^2 \phi_1 } +
    \frac{\,(l_{1}+m)^2-1/4}{{\sin}^2 \phi_1 }
    \label{hier}
\eea
They originate the one-dimensional Hamiltonian hierarchy (\ref{hier}), starting from $H_{(0)}^{\phi_1}$. The Hamiltonians  $H_{(m)}^{\phi_1}$ satisfy the
fundamental relation
\be\label{co}
H_{(m)}^{\phi_1} = A^{+}_{m}\,A^{-}_{m} + \lambda_{m} =
A^{-}_{m-1}\,A^{+}_{m-1} +
    \lambda_{m-1}
\ee
so that $A_m^\pm$ are shape invariant intertwining operators, i.e. 
\be\label{ah}
A^{-}_{m}H_{(m)}^{\phi_1}=H_{(m+1)}^{\phi_1}A^{-}_{m}\ ,\qquad
A^{+}_{m}H_{(m+1)}^{\phi_1}=H_{(m)}^{\phi_1}A^{+}_{m}
\ee
Hence, from a formal point of view, the operators $A^\pm_{m}$ acting on a Hamiltonian eigenfunction
will give another eigenfunction of a consecutive Hamiltonian in the hierarchy with the
same eigenvalue.
If we design the eigenfunction spaces of $H_{(m)}^{\phi_1}$ (as differential operators) by ${\cal H}_{m}^{\phi_1}$, then we have
$$
A^{-}_{m} : {\cal H}_{m}^{\phi_1} \to {\cal H}_{m+1}^{\phi_1},\qquad
A^{+}_{m} : {\cal H}_{m+1}^{\phi_1} \to {\cal H}_{m}^{\phi_1}
$$
In principle, the discrete spectrum and the physical eigenstates of $H_{(0)}^{\phi_1}$  could be obtained from the fundamental states $f_{(m)}^0$  and their eigenvalues of all the Hamiltonians in the hierarchy $\{H_{(m)}^{\phi_1}\}$.  These fundamental states are determined by $A^{-}_{m}\,f^{0}_{(m)} = 0$, giving the solutions (up to a normalization
constant)
\[
    f^{0}_{(m)}(\phi_1)=  \cos^{l_0 + m+1/2}\phi_1
    \sin^{l_1+m+1/2}\phi_1
\]
with eigenvalues $\lambda_{m} = (l_0 + l_1 + 2\,m+1)^2$.
Often the intertwining operators (\ref{aa}) are written in the form
\be\label{sp1}
A_m^\pm =
\pm\p_{\phi_1} + \om_m(\phi_1),\qquad
\om_m(\phi_1)= \frac{\p_{\phi_1}
f^{0}_{(m)}(\phi_1)}{f^{0}_{(m)}(\phi_1)}
\ee
where $\om_m(\phi_1)$ is called superpotential function.

In order to go from the ground eigenstate, $f_{(m)}^0$ of
$H_{(m)}^{\phi_1}$, up to the excited eigenfunction, $f_{(0)}^{m}$ of
$H_{(0)}^{\phi_1}$, with the same eigenvalue, we apply consecutive operators $A^+$
\be\label{excited}
    f_{(0)}^{m} = A^{+}_{0}\,A^{+}_{1}\cdots
A^{+}_{m-1}\,f_{(m)}^0
\ee
obtaining explicitly
\be \label{Jacob}
      f_{(0)}^{m}=\,N\,
     \sin^{l_1+1/2}\phi_1\,\,\cos^{l_0+1/2}\phi_1\,\,
       P_{m}^{(l_1,l_0)}(\cos(2\,\phi_1))
\ee
where $P_{n}^{(a,b)}(x)$  are Jacobi polynomials and $N$ a
normalization constant. Therefore, the spectrum of the first separating Hamiltonian (\ref{sisn21}) is
given by
\be\label{aespectro}
\a = \lambda_{m}=(l_0 + l_1 + 2\,m+1)^2,\qquad m\in\Z^+
\ee

The following two subsections are devoted to characterise the Lie algebras of shape invariant intertwining operators for the one-dimensional Hamiltonian hierarchies. They will constitute a useful pattern for the two-dimensional Hamiltonians of section~\ref{dynamicalsymmetries}.


\subsect{The dynamical algebra $u(2)$} \label{theu(2)dynamicalalgebra}

Starting from the operators $A^\pm_{m}$ let us define free-index
operators $A^\pm$ acting inside the total space $\oplus_{m}{\cal H}_{m}$, in the following way \cite{fernandez,refined00}:
\be \label{aas}
\begin{array}{l}
A^+ f_{(m+1)}:= \frac12 A^+_{m} f_{(m+1)}\propto \tilde f_{(m)}
\\[0.30cm]
A^- f_{(m)}:= \frac12 A^-_{m} f_{(m)} \propto \tilde f_{(m+1)}
\\[0.3cm]
A \,f_{(m)}:= -\frac12(l_0 {+} l_1 {+} 2 m )f_{(m)}\propto
f_{(m)}
\end{array}
\ee
where $f_{(m)}$ (or $\tilde f_{(m)}$) denotes an eigenfunction of $H_{(m)}$.
This action can be extended to linear
combinations of eigenfunctions by linearity. With this
convention we can rewrite (\ref{co}) and (\ref{aas}) simply as the commutators
\be \label{com}
[A,A^{\pm}] = \pm A^{\pm},\qquad
[A^-,A^+] = - 2 A
\ee
assuming that the action is on any (linear combination of) $f_{(m)}$'s. The commutators (\ref{com})
close  the Lie algebra $su(2)$, whose Casimir element is given by
${\cal C}= A^+A^- +A(A-1)$. The eigenvalues of $\cal C$, labeling the
 irreducible unitary representations (IUR), are $j(j+1)$, where $2j\in \Z^+$. The
dimension of the support spaces of these IUR's is, obviously,  $2j+1$. We make use of the standard
notation $|j,s\rangle$ for an
$A$-eigenvector with  eigenvalue $s$, inside the `$j$-representation'.

Now, we can identify the eigenstates of the Hamiltonians $H^{\phi_1}_{(m)}$ in terms of
representation vectors $|j,s\rangle$. First, let us consider the ground states
$f_{(m)}^0$ characterized by
\be\label{fu}
A^-f_{(m)}^0=0,\qquad
A\,f_{(m)}^0=-[(l_0+l_1+2m)/2] f_{(m)}^0
\ee
following the notation
(\ref{aas}). These relations suggest the identification (up to a normalization
constant)
\[
f_{(m)}^0 = |j_m,-j_m\rangle, \qquad j_m=(l_0+l_1+2m)/2
\]
To see
that indeed this is the case we need to define the whole representation space as well as an
inner product.
Thus,  consider the space ${\cal L}^2[0,\pi/2]$ of  square integrable functions in the interval
$[0,\pi/2]$. Then, the wavefunctions obtained from the ground state
$f_{(m)}^0$ by the consecutive action of the operator $A^+$ will span the representation space of a $j_m$-representation, with
$j_m=(l_0+l_1+2m)/2$, provided that both $l_0+m$ and $l_1+m$ belong to  $\Z^+$.  The wavefunctions of the space so generated vanish
at the end points  (i.e. $\Psi(0)=\Psi(\pi/2)=0$), and the hermiticity relations
$(A^-)^{\dagger}=A^+$, $A^{\dagger} = A$ are implemented in all the space.
Hence, under these conditions, $(A^+)^k f_{(m)}^0$ can be identified, up to
normalization, with the vector state
$|j_m,-j_m+k\rangle$.

As a consequence, the excited states obtained in this way for any Hamiltonian in
a factorization hierarchy where $l_0$ and $l_1$ are positive integers, correspond to
IUR-vector states. For instance, the eigenstate of the $k$-th excited level of
$H^{\phi_1}_{(0)}$ is
\[
f_{(0)}^k\equiv |j_k+k,-j_k+k\rangle,\qquad j_k=(l_0+l_1+2k)/2, \qquad k=0,1,2\dots
\]
and $H^{\phi_1}_{(0)}$ (as well as any $H^{\phi_1}_{(m)}$) can be expressed in 
terms of the $su(2)$-Casimir
$\cal C$ acting on such representations
\[\label{ca}
H^{\phi_1}_{(0)}=4({\cal C} +1/4)
\]
Therefore, the eigenvalue equation for any of the excited states can be written
as follows
\[\begin {array}{lll}
H^{\phi_1}_{(0)} f_{(0)}^k &\equiv & 4({\cal C}+1/4)|j_0+k,-j_0+k\rangle \\[1.ex]
&= & 4(j_0+k+1/2)^2
|j_0+k,-j_0+k\rangle =(l_0+l_1 +2k+1)^2 f_{(0)}^k
\end{array}\]
with $k=0,1,2,\dots$.

It will be convenient to consider a new diagonal operator $D$, to
be added to the generators of  $su(2)$  (\ref{aas}), 
define by
\[\label{d}
D f_{(m)}:= (l_0-l_1) f_{(m)}
\]
It is immediate to see that $D$ commutes with any other operator of  $su(2)$ giving rise to
the Lie algebra $u(2)$.
In this way any eigenstate in the Hamiltonian hierarchy can be
characterised completely by an eigenfunction of an $u(2)$-IUR. Without
$D$ we would have an ambiguity due to the fact that different fundamental states
with values of $l_0$ and $l_1$  giving the same $j_0=(l_0+l_1)/2$ would lead to the
same $j$-representation of $su(2)$.

It is worthy noting that when
$l_0$ or $l_1$ are not in $\Z^+$ the eigenfunctions
and spectrum of the Hamiltonian hierarchies are still given by (\ref{Jacob}) and (\ref{aespectro}),
but, these states belong to non-unitary
representations of $u(2)$.

\subsect{The dynamical algebra $so(4)$} \label{theso(4)dynamicalalgebra}

As we have just seen in the previous subsection the eigenstates sharing
the same energy of the one-dimensional Hamiltonian hierarchies in the
variable $\phi_1$ are given in terms of IUR's of the  dynamical
algebra $u(2)$. However, in this respect, there is a point not quite
satisfactory: different $u(2)$-IUR's may correspond to states with the
same energy. We would prefer a larger dynamical algebra with a simpler
correspondence, i.e., such that only one of its IUR's gives all the
eigenstates with the same energy in the hierarchy.

In order to build up a  dynamical algebra having these properties, let us introduce the two-dimensional parameter space $(l_0,l_1)$. Any operator with one subindex defined in subsections~\ref{thephi1--factorization} and~\ref{theu(2)dynamicalalgebra} will change to a two-subindex notation in the following  way:
\begin{enumerate}
\item 
 The one-dimensional Hamiltonian (\ref{sisn21}) will be denoted by $H_{(l_0,l_1)}$
\[\label{h01}
H_{(l_0,l_1)}^{\phi_1}=  -\d(\phi_1)^2
    +\frac{ l_{0}^2-1/4}{{\cos}^2 \phi_1 } +
    \frac{l_{1}^2-1/4}{{\sin}^2 \phi_1 }
\]
Its eigenfunctions will be designed by $f_{(l_0,l_1)}$.

\item
The factor operators $A_0^\pm$ in (\ref{aa}) will be rewritten as $A_{(l_0,l_1)}^\pm$
\[\label{a01}
A_{(l_0,l_1)}^{\pm} = \pm \d(\phi_1) -  (l_0+1/2) \,\tan \phi_1 +
       (l_1+1/2)\,\cot \phi_1,\quad
A_{(l_0,l_1)}= -\frac12(l_0 {+} l_1  )
\]
\end{enumerate}

Now, in this way, relations (\ref{ah}) can be expressed as
\be\label{ahh}
A^{-}_{(l_0,l_1)}H_{(l_0,l_1)}^{\phi_1}=
H_{{(l_0+1,l_1+1)}}^{\phi_1}A^{-}_{(l_0,l_1)}\ ,\qquad
A^{+}_{(l_0,l_1)}H_{{(l_0+1,l_1+1)}}^{\phi_1}=
H_{{(l_0,l_1)}}^{\phi_1}A^{+}_{(l_0,l_1)}
\ee
With this convention we can also define  the free-subindex operators
$A^\pm,A, D$ as in (\ref{aas}).

On the other hand, notice that each two-parameter Hamiltonian $H_{(l_0,l_1)}^{\phi_1}$ is invariant under the reflections
\[
I_0:(l_0,l_1)\to(-l_0,l_1),\qquad
I_1:(l_0,l_1)\to(l_0,-l_1)
\]
This property gives rise to a second
factorisation (see also \cite{barut,quesne,dutt})
via conjugation of the operators of the first factorisation by the reflection opera tors
\[
\begin{array}{lll}
I_0 A^\pm I_0=\tilde A^\pm,
\qquad  &I_0 A I_0=\tilde A,
\qquad  &I_0 D I_0=\tilde D\\[0.3cm]
\tilde I_1 A^\pm I_1=\tilde A^\mp,
\qquad  &I_1 A I_1=-\tilde A,
\qquad  &I_1 D I_1=-\tilde D
\end{array}
\]
Explicitly
\be\label{ta01}
\tilde A_{(l_0,l_1)}^{\pm} =
\pm \d(\phi_1) +  (l_0{-}1/2) \,\tan \phi_1 +
       (l_1+1/2)\,\cot \phi_1,\quad
\tilde A_{(l_0,l_1)}= -\frac12(-l_0 {+} l_1 )
\ee
The above operators $\{ \tilde A, \tilde A^{\pm}\}$ generate a  Lie algebra isomorphic to $su(2)$ denoted by 
$\widetilde{su}(2)$. Since $su(2)$ and $\widetilde{su}(2)$ commute and,
essentially, $D$ and
$\tilde D$ coincide with $\tilde A$ and $A$, respectively, the complete
dynamical algebra has the structure of a direct sum $su(2)\oplus
\widetilde{su}(2)\approx so(4)$.

If we allow to act with the $so(4)$ generators on an Hamiltonian
$H_{(l_0,l_1)}$
we will get a two-dimensional parameter lattice of Hamiltonians which
constitute a $so(4)$-hierarchy fixed by the initial values
$(l_0,l_1)$: $\{ H_{l_0-n+m,l_1+n+m}\}$, $m,n\in\Z$. Each energy level
of this Hamiltonian hierarchy is degenerated and the eigenstates belong to $so(4)$-representations. 

Let us concentrate on the hierarchies associated to IUR's of $so(4)$. Now, these  $so(4)$-IUR's are fixed by
the fundamental (or lowest weight) states satisfying
\be\label{fuu}
A^-_{(l_0,l_1)}f_{(l_0,l_1)}^0=
\tilde A^-_{(l_0,l_1)}f_{(l_0,l_1)}^0=0
\ee
These are realized, up to a constant, by the wavefunctions
\be\label{o}
f_{(0,n)}^0= \cos^{1/2}\phi_1\sin^{n+1/2}\phi_1,\qquad  n\in \Z^+
\ee
where we have taken  $l_0=0$ and $l_1=n$.
We see also that the state (\ref{o}) is stable under $I_0$ (i.e. 
$I_0 f_{(l_0,l_1)}^0 = f_{(l_0,l_1)}^0$), and comes into the
other fundamental state (annihilated by $A^+$ and $\tilde A^+$: the highest weight) of the
same representation. Hence,  these representations will be invariant
under $I_0$ and $I_1$.
Therefore, the $so(4)$-IUR's obtained from (\ref{o}) are symmetric 
tensor products that can be denoted by
\[ \label{rep}
j\otimes j,  \qquad\qquad j=l_1/2= n/2,\qquad n\in\Z^{\geq 0}
\]
where `$j$' stands  for a $j$-representation of $su(2)$.
In this way  the degenerancy of the $n$-th energy level is
$(n+1)\times(n+1)$, which is composed of $n+1$ IUR's of  $u(2)$ each of them of dimension $n+1$.  

The Hamiltonians in this hierarchy can be expressed in terms
of any of the $su(2) ($or  $\widetilde{su}(2)$) Casimir operators
$H_{(l_0,l_1)}=4({\cal C} +1/4)=4(\tilde{\cal C} +1/4)$.
With the help of all the discrete reflections we get directly its
expression also in terms of the $so(4)$-Casimir
\[\label{ca}
\begin{array}{ll}
H_{(l_0,l_1)}&=\, ({\cal C} +1/4)+I_0({\cal C} +1/4)I_0
+I_1({\cal C} +1/4)I_1
+I_0 I_1({\cal C} +1/4)I_0 I_1 \\[2.ex]
&=\,  \{A^+,A^-\}+2 A^2+ \{\tilde A^+,\tilde A^-\}+2 \tilde A^2 +1\\[2.ex]
&=\,  \{A^+,A^-\}+ \{\tilde A^+,\tilde A^-\}+ {L_0}^2 + {L_1}^2 +1
\end{array}
\]
where the diagonal operators $L_0$ and $L_1$ are defined by
\[
L_0\, f_{(l_0,l_1)} = l_0\,f_{(l_0,l_1)},\qquad
L_1\, f_{(l_0,l_1)} = l_1\,f_{(l_0,l_1)}
\]
Certainly, some $so(4)$-hierarchies (those
corresponding to the IUR's previously described) may have Hamiltonians
whose explicit expressions coincide
$$
H_{(l_0,l_1)}=H_{(-l_0+1,l_1)}=H_{(l_0,-l_1+1)}
$$
 and the same happens
with their corresponding eigenstates. But we can not get rid of this
multiplicity unless we enlarge the ambient space. 

Another natural question is
whether there are other intertwining shape-invariant operators inside
the $so(4)$-hierarchy. We can build, for instance, other pairs of
operators through the composition of those already known
\[
{\cal X}^\pm = A^\pm \tilde A^\pm,\qquad
{\cal Y}^\pm = A^\pm \tilde A^\mp
\]
This kind of shape-invariant operators change two units either the
parameter $l_0$ or $l_1$ (but not both at the same time). When we restrict to $l_0=0$ or $l_1=0$ there are also first order intertwining operators changing one unit the
nonvanishing parameter. This feature is not so special; it is also
shared by the `radial oscilator' hierarchies  \cite{confluent} (which are closely related
to the ones presented  here).

For other Hamiltonian $so(4)$-hierarchies the physical eigenstates are
described by non-unitary representations that are not invariant under
both reflections. In this respect, their description becomes more
involved, so that one must be very careful in these cases.


\subsect{The $\phi_2$--factorization} \label{thephi2--factorization}

Now, let us return to the separation process started in subsection~\ref{thephi1--factorization}.
The second equation (\ref{sisn22}) obtained from the initial
separation of variables can be dealt with along the same lines,
substituting the eigenvalues  obtained from the previous factorization,
$\a = \l_{m} = (l_0 + l_1 +2m)^2$. The
most relevant fact, here, is that the new factorization leads to a
degeneration of the energy levels which suggest that the underlying
dynamical symmetry could be larger, as it will be confirmed in the next
section. Thus, substituting in (\ref{sisn22}), we have
\be\label{2}
\begin{array}{ll}
H_{(0)}^{\phi_2} &=
{\displaystyle -\p_{\phi_2}^2 +  \tan(\phi_2)\p_{\phi_2} +
\frac{ (l_0 + l_1 +2m+1)^2}{ \cos^2(\phi_2) }
+\frac{l_2^2-1/4)}{\sin^2(\phi_2)}}\\[0.35cm]
&=\left\{ \p_{\phi_2} - (l_0 + l_1 +2(m+1)) \tan(\phi_2) + (l_2+1/2)\,
\cot(\phi_2)\right\}\\ [0.2cm]
&\quad\quad \times \left\{ -\p_{\phi_2} -  (l_0 + l_1 +2m+1)
\tan(\phi_2) + (l_2+1/2)\, \cot(\phi_2)\right\}\\[0.2cm]
&\qquad \qquad +\;  (l_2+l_0 + l_1 +2m+3/2)(l_2+(l_0 + l_1 +2m+5/2)\\[0.3cm]
&\equiv M^+_0 M^-_0 + \mu_0.
\end{array}
\ee
This is the first one of the Hamiltonian hierarchy $H_{(n)}^{\phi_2}$ in the
variable ${\phi_2}$,
$$
H_{(n)}^{\phi_2} = M^+_n M^-_n + \mu_n = M^-_{n-1} M^+_{n-1} +
\mu_{n-1}
$$
where
\[\begin{array}{l}
M_n^\pm= \pm \p_{\phi_2} -
 (l_0 + l_1 +2(m+1) +n) \tan(\phi_2) + (l_2+n+1/2)\cot(\phi_2)\\[0.3cm]
\mu_n =  (l_1 +l_0 +l_2 + 2n + 2m +3/2)(l_2+l_1 +l_0 +2n +2m+5/2)
\end{array}\]
Now, the values for the energy (following closely the same
arguments of section~\ref{thephi1--factorization}) are given by
 \be\label{energia} 
 E = \mu_n =
(l_1 +l_0 +l_2 + 2n + 2m +3/2)(l_2+l_1 +l_0 +2n +2m+5/2)
\ee 
The fundamental states $g_{(n)}^{0}$ for this factorization are 
\[
g_{(n)}^{0}(\phi_2) = N \cos^{l_1 +l_0 \phi_2+2m+1 } \phi_2\; 
\sin^{l_2 +n+1/2 }\phi_2
\]
 and the eigenfunctions $g_{(0)}^{n}$ of the
initial Hamiltonian (\ref{2}) can be written in the form
\be \label{ef2} 
g_{(0)}^{n}(\phi_2) = \cos^{l_1+l_0+2m+1} \phi_2\;
\sin^{l_2+1/2} \phi_2\;P_n^{(l_2+1/2,l_1 +l_0 +2m+1)}(\cos 2\phi_2).
\ee
The commutation relation for the relevant free-index operators
$M^\pm$, defined in a similar way as $A^\pm$ in (\ref{aas}), is
again that of   $su(2)$, 
\[\label{bes} 
 [M^-,M^+] = -4 (l_1+l_0
+l_2 +2m+2n + 1)\equiv -2 M 
\]
The eigenfunctions
(\ref{ef2}) are square-integrable, but the representations are
unitary provided that, besides the previous conditions on $l_0$ and
$l_1$,  the parameter $l_2$  be  also a positive integer number.

In summary, if we finally join the results of both factorizations, the
square-integrable eigenfunctions  of the Hamiltonian (\ref{h})   in the separable variables
$(\phi_1,\phi_2)$ are given by the products
\be \label{gs}
\Phi_{m,n}(\phi_1,\phi_2) = f^m_{(0)}(\phi_1)\,
g^n_{(0)}(\phi_2),\qquad  m,n\in \Z^+
\ee
where the components have the polynomial expressions (\ref{Jacob})
and (\ref{ef2}). The corresponding eigenvalues given
in (\ref{energia}) are degenerated for those  values of $m$ and $n$ whose sum $m+n$
keeps constant (see also, for instance, Ref.~\cite{pogosyan}).

\sect{Dynamical symmetries} \label{dynamicalsymmetries}

The spectrum obtained by the methods of section~\ref{su3hamiltoniansystem} suggest
 the existence of a bigger dynamical algebra of the Hamiltonian hierarchy. This is the point 
 that we want to
address here developing exhaustively the concept of intertwining (shape
invariant) operators for this kind of Hamiltonians. Such operators will supply
us with a more consistent picture of the spectrum and eigenfunctions. Thus, based on the considerations of subsections~\ref{theu(2)dynamicalalgebra} and~\ref{theso(4)dynamicalalgebra}, we will
introduce three sets of intertwining operators closing the  Lie algebra $u(3)$. Then, in the following subsection, we will enlarge this algebra to $so(6)$ by means of the relevant reflections.

\subsect{The Hamiltonian $u(3)$--hierarchies}

\subsubsect{The set $\bf \{A^+, A^-, A\}$}

As we will use some properties of section~\ref{su3hamiltoniansystem} in a different direction,
it is convenient to introduce another notation more appropriate to rewrite
some previous results. The Hamiltonian (\ref{hh}) characterized by the
parameters
$\ell\equiv (l_0,l_1,l_2)$ will be referred to as
$H_{(l_0,l_1,l_2)}$, and the operators defined by (\ref{aa})  will be
taken henceforth  with a three-fold subindex
\be\label{als}
A_{(l_0,l_1,l_2)}^{\pm}=
\pm
\d(\phi_1) -  (l_0+1/2) \,\tan \phi_1 + (l_1+1/2)\,\cot \phi_1
\ee
Since the
differential operators (\ref{aa}) and (\ref{als}) depend only on the variable
$\phi_1$,  they do not affect the part in the total Hamiltonian (\ref{hh})
depending on the second separable variable $\phi_2$. So that, in the same
way as (\ref{ahh}) we have the intertwining relations
$$\begin{array}{l}
A^{-}_{(l_0,l_1,l_2)}H_{(l_0,l_1,l_2)}=
H_{(l_0+1,l_1+1,l_2)}A^{-}_{(l_0,l_1,l_2)}\\[0.3cm]
A^{+}_{(l_0,l_1,l_2)}H_{(l_0+1,l_1+1,l_2)}=H_{(l_0,l_1,l_2)}A^{+}_{(l_0,l_1,l_2)}
\end{array}$$
This means that now $A_{(l_0,l_1,l_2)}^{-}$ is acting on  eigenstates of
$H_{(l_0,l_1,l_2)}$ leading to eigenstates of $H_{(l_0+1,l_1+1,l_2)}$, while
$A_{(l_0,l_1,l_2)}^{+}$ does it in the opposite way (later we will comment on
the square-integrability conditions through unitary representations).

If we include the normalizing constant just as in (\ref{aas}), and define global
operators acting on eigenfunctions of this class of Hamiltonians in the form
\[\begin{array}{l}\label{mas}
A^+ \Phi_{(l_0+1,l_1+1,l_2)}:= \frac12 A^+_{(l_0,l_1,l_2)}
\Phi_{(l_0+1,l_1+1,l_2)} \propto \tilde \Phi_{(l_0,l_1,l_2)}\\[0.3cm]
A^- \Phi_{(l_0,l_1,l_2)}:= \frac12 A^-_{(l_0,l_1,l_2)}
\Phi_{(l_0,l_1,l_2)} \propto \tilde \Phi_{(l_0+1,l_1+1,l_2)}\\[0.3cm]
A\, \Phi_{(l_0,l_1,l_2)}:= -\frac12(l_0 + l_1)\Phi_{(l_0,l_1,l_2)}
\end{array}\]
we are lead to the standard $su(2)$ commutators (\ref{com}). Here, we want
to stress again that now these operators are acting on the total wavefunction of
complete Hamiltonians like $H_{(l_0,l_1,l_2)}$, not just on a factor function
in only one variable.

In order to introduce other sets of operators we will use the fact that the
Hamiltonian (\ref{hh}) can be separated in other coordinate systems. Since the
axes $(s_0,s_1,s_2)$ play a symmetric role in the Hamiltonian, we will take their cyclic
rotations to get two other  sets of coordinates and, hence, new sets of  intertwining operators.

\subsubsect{The set $\bf \{B^+, B^-, B\}$}
We will take the spherical coordinates choosing as third axis not $s_2$, but
$s_1$, i.e.
\be\label{coo2}
   s_2 = \cos\xi_2\,\cos\xi_1,\qquad
   s_0 = \cos\xi_2\, \sin\xi_1,\qquad
   s_1 \,= \sin\xi_2\,
\ee
Then, the initial Hamiltonian is also separated in the coordinates $(\xi_1,\xi_2)$. In
particular, we can build the operators $B_{(l_0,l_1,l_2)}^{\pm}$ in a similar way
as $A_{(l_0,l_1,l_2)}^{\pm}$. From the coordinate systems (\ref{coo}) and (\ref{coo2}) we  easily 
arrive at the following expressions for the new set in terms of the
initial coordinates $(\phi_1, \phi_2)$
\be\label{bls}
B_{(l_0,l_1,l_2)}^{\pm}=
\pm
(\sin\phi_1\tan \phi_2\d(\phi_1) +\cos\phi_1\d(\phi_2)) -  (l_2{+}1/2) \,\cos\phi_1{\rm
cot}\phi_2 + (l_0{+}1/2)\,{\rm sec}\phi_1{\rm tan}\phi_2
\ee
These operators intertwin the pair of Hamiltonians
\[\begin{array}{l}
B^{-}_{(l_0,l_1,l_2)}H_{(l_0,l_1,l_2)}=
H_{(l_0+1,l_1,l_2+1)}B^{-}_{(l_0,l_1,l_2)}\\[0.3cm]
B^{+}_{(l_0,l_1,l_2)}H_{(l_0+1,l_1,l_2+1)}=H_{(l_0,l_1,l_2)}B^{+}_{(l_0,l_1,l_2)}
\end{array}\]
The `global' operators, defined by
\[ \begin{array}{l}\label{mas2}
B^+ \Phi_{(l_0+1,l_1,l_2+1)}:= \frac12 B^+_{(l_0,l_1,l_2)}
\Phi_{(l_0+1,l_1,l_2+1)} \propto \tilde \Phi_{(l_0,l_1,l_2)}\\[0.3cm]
B^- \Phi_{(l_0,l_1,l_2)}:= \frac12 B^-_{(l_0,l_1,l_2)}
\Phi_{(l_0,l_1,l_2)} \propto \tilde\Phi_{(l_0+1,l_1,l_2+1)}\\[0.3cm]
B \Phi_{(l_0,l_1,l_2)}:= -\frac12(l_0 + l_2 )\Phi_{(l_0,l_1,l_2)}
\end{array}\]
 also close a new  $su(2)$.

\subsubsect{The set $\bf \{C^+, C^-, C\}$}

Finally, taking the spherical coordinates around the $s_0$ axis,
\[\label{coo3}
   s_1 = \cos \theta_2 \,\cos \theta_1 ,\qquad
   s_2 = \cos \theta_2 \, \sin \theta_1 ,\qquad
   s_0 \,= \sin\theta_2
\]
the Hamiltonian is also separated in the variables $\{\theta_1,\theta_2\}$ and we get 
a new pair of operators, that written in terms of the initial  $\phi_1$ and $\phi_2$
variable, take the expression 
\be\label{cls}
C_{(l_0,l_1,l_2)}^{\pm}=
\pm
(\cos\phi_1\tan \phi_2\d(\phi_1) -\sin\phi_1\d(\phi_2)) +  (l_1{-}1/2)
\,{\rm cosec}\phi_1{\rm tan}\phi_2 + (l_2{+}1/2)\,{\rm sin}\phi_1{\rm cot}\phi_2
\ee
These operators act as intertwiners of the Hamiltonians in the following way
\[\begin{array}{l}
C^{-}_{(l_0,l_1,l_2)}H_{(l_0,l_1,l_2)}=
H_{(l_0,l_1-1,l_2+1)}C^{-}_{(l_0,l_1,l_2)}\\[0.3cm]
C^{+}_{(l_0,l_1,l_2)}H_{(l_0,l_1-1,l_2+1)}=H_{(l_0,l_1,l_2)}C^{+}_{(l_0,l_1,l_2)}
\end{array}\]
The `global' operators are defined by
\[\begin{array}{l}\label{mas3}
C^+ \Phi_{(l_0,l_1-1,l_2+1)}:= \frac12 C^+_{(l_0,l_1,l_2)}
\Phi_{(l_0,l_1-1,l_2+1)} \propto  \tilde \Phi_{(l_0,l_1,l_2)}\\[0.3cm]
C^- \Phi_{(l_0,l_1,l_2)}:= \frac12 C^-_{(l_0,l_1,l_2)}
\Phi_{(l_0,l_1,l_2)} \propto \tilde \Phi_{(l_0,l_1-1,l_2+1)}\\[0.3cm]
C\, \Phi_{(l_0,l_1,l_2)}:= -\frac12(-l_1 + l_2 )\Phi_{(l_0,l_1,l_2)},
\end{array}\]
closing the third algebra  $su(2)$. Notice that $C=B-A$.

In fact, as we saw in section~\ref{su3hamiltoniansystem}, each separable system gives rise to two
sets of intertwining operators (in that section distinguised by means
of the tilde). However, here we have made a `good' choice of the above
three sets that will close a Lie algebra (on this point see section~\ref{theso(6)hierarchy}).

\subsubsect{The complete  algebra $\bf u(3)$}

Now, we can join all the transformations
 above defined, $A^{\pm}, A, B^{\pm}, B, C^{\pm}, C$, and commute any two
of them to check that indeed they close a Lie algebra  $su(3)$. The  nonvanishing commutators are 
\[\begin{array}{llll}
[A^3,A^{\pm}] = \pm A^{\pm} \; & [ A^-,A^+]= 2 A\;  & [A^+,B^-]
=C^- \; &[A^+,B] = -A^+/2 \\ [1.ex]
[A^+,C^+]= -B^+\; & [A^+,C] = A^+/2\;  & [A^-,B^+] =-C^+\;  & [A^-,B]= A^-/2\\ [1.ex]
[A^+,C^+] = B^-\;  &[A^-,C] = -A^-/2\;  &[A,B^+]= B^+/2\;
&[A,C^+] =-C^+/2 \\[1.ex]
[A,C^-]=C^-/2\;   &[B,B^\pm] = \pm B^\pm\; &[B^-,B^+] = 2 B\;
&[B^+,C^-]=-A^+ \\[1.ex]
[B^+,C]= -B^+/2\; &[B^-,C^+]=C^+/2   \; &[B^-,C]=B^-/2\;
&[B,C^+] = C^+/2\\ [1.ex]
[B,C^-] =C^-/2\;  & [C,C^\pm] = \pm C^\pm\; &[C^+,C^-] = 2C \; &[A^-,C^-] =B^-
\end{array}\] 
The Casimir operator is given by
\be\label{cas}
 {\cal C}= A^+A^- + B^+B^- + C^+C^- + \frac23 A(A-3/2)
+ \frac23 B(B-3/2) + \frac23 C(C-3/2)
 \ee
  In order to complete an algebra
$u(3)$  we can add a diagonal operator $D$ commuting with all
the above transformations.  It is a central operator, i.e.
\[
D:=l_0-l_1-l_2,\qquad [D,\cdot] = 0
\] 
 We can also adopt the global
operator convention $H$ for the Hamiltonians in the hierarchy by
defining its action on the eigenfunctions $\Phi_{(l_1,l_2,l_3)}$ of
$H_{(l_1,l_2,l_3)}$ by 
\[ 
H\Phi_{(l_1,l_2,l_3)}:= H_{(l_1,l_2,l_3)}
\Phi_{(l_1,l_2,l_3)}
 \] 
In this way we can express the Hamiltonian
$H$ in terms of both operators ${\cal C}$ and $D$
\be \label{esp} 
H= 4\,{\cal C}-\frac13\, D^2+\frac{15}4 
\ee 
In the case of one-dimensional
systems, one (first order) intertwining set $\{A^\pm\}$ for the
Hamiltonian gives rise to its factorization. However, for
Hamiltonians with more degrees of freedom (more components, or in
more dimensions) the relationship of $H$ with these operators, in
general, turns out to be more complex. In our case the set
$\{A^\pm,B^\pm,C^\pm\}$ according to expressions (\ref{cas}) and (\ref{esp}) is
enough to express the Hamiltonian as a certain quadratic function
$H= h(A^+A^-, B^+B^- , C^+C^-)$ generalizing the usual
factorization.

In summary, we have built an algebra $u(3)$ of intertwining operators that, once fixed the initial Hamiltonian with parameter values
$(l_0,l_1,l_2)$,
gives rise to a two-parameter Hamiltonian hierarchy
\[
 \{H_{(l_0+m,l_1+m-n,l_2+n)} \},\qquad
m,n\in\Z
\]
where the points $(l_0+m,l_1+m-n,l_2+n)$ lie on a certain plane $D=d_0$.
In this subsection we will consider this special hierarchy, together with its eigenstates,  connected to the IUR's of $u(3)$. The states of such representations are square integrable and, therefore,
should take part of the physical eigenfunctions whose energy eingenvalues belong
to the spectrum.

In order to build an IUR we start from a fundamental state $\Phi$
annihilated by
$A^-$ and $C^-$ (two simple roots of $su(3)$)
\be \label{ac}
A_\ell^-\Phi_\ell = C_\ell^-\Phi_\ell =0
\ee
with $\ell=(l_0,l_1,l_2)$.  Such states exist only when $l_1=0$, taking the explicit form
\be\label{ground}
\Phi_\ell(\phi_1,\phi_2)=
N \,\cos^{l_0{+}1/2}\phi_1\;\sin^{1/2}\phi_1\;\cos^{l_0+1}\phi_2\;\sin^{l_2{+}1/2}\phi_2
\ee
where $N$ is a normalizing constant. The diagonal operators $A$ and $C$ act on $\Phi_\ell$
as
\be\label{mn}
\begin{array}{lll}
A\, \Phi_\ell = -l_0/2\, \Phi_\ell,\quad & l_0 = m,\qquad l_1=0,\qquad & m=0,1,2,\dots\\[0.3cm]
C\, \Phi_\ell = -l_2/2\, \Phi_\ell,\quad & l_2 = n,\qquad & n=0,1,2,\dots
\end{array}
\ee
This means that $\Phi_\ell$ is a fundamental state of the representations
$j_1=m/2$ of the  subalgebra $su(2)$  generated by $\{A^\pm,A\}$, and $j_2=n/2$ of the corresponding $su(2)$ determined by $\{C^\pm,C\}$. Such a representation of $su(3)$ will be denoted 
$(m,n)$ with $m,n\in\Z^+$. The points labeling
the states of this   representation obtained from $\Phi_\ell$  lie on
the plane $D=m-n$ inside the $\ell$-parameter space.

The energy for the states of the IUR's determined by the fundamental state
(\ref{mn}) with  the parameters $(l_0,0,l_2)$, according to (\ref{esp}) is given by
\be\label{ee}
E=(l_0 + l_2 + 3/2) ( l_0  + l_2 + 5/2)=(m + n + 3/2) ( m  + n + 5/2)
\ee
Therefore, the IUR's fixed by $(m,n)$ with the same value $m+n$ will lead to states with the same energy. We call such IUR's an iso-energy series and they will be examined under the light of the algebra  $so(6)$ in the following section.
The values for the energy (\ref{ee})
coincide with the ones computed by the method of variable separation of 
section~\ref{su3hamiltoniansystem}, as can be seen from (\ref{energia}) once the replacement $l_1=1/2$ is performed. 
We can  also check that in this case the ground state (\ref{gs}) coincides with those fixing 
an IUR (\ref{ground}). 
\subsect{The $so(6)$--hierarchy} \label{theso(6)hierarchy}

Following the pattern and motivation of section~\ref{theso(4)dynamicalalgebra}, we will consider the relevant discrete symmetries in order to find a larger dynamical algebra.

It is obvious that the Hamiltonian $H_{(l_0,l_1,l_2)}$ is invariant under reflections in the parameter space $\{(l_0,l_1,l_2)\}$
\[
I_0:(l_0,l_1,l_2)\to (-l_0,l_1,l_2),\qquad
I_1:(l_0,l_1,l_2)\to (l_0,-l_1,l_2),\qquad
I_2:(l_0,l_1,l_2)\to (l_0,l_1,-l_2)
\]
Each of these symmetries can be directly implemented in the eigenfunction space, leading through  conjugation to another set of
intertwining operators that close a Lie algebra isomorphic to  $u(3)$ and denoted by ${}_iu(3)$
\[
{}_i\!X = I_i \, X\,  I_i,\qquad\qquad X\in u(3),\quad {}_i\!X\in {}_i\!u(3),\quad i=0,1,2
\]
The intertwining operators of ${}_iu(3)$ connect eigenstates of Hamiltonians
whose parameters $(l_0,l_1,l_2)$ belong to the planes ${}_iD = k_i$, being $k_i$ certain real constants. We will choose the following convention for the resulting generators
\[
\begin{array}{ll}
\{A^\pm,B^\pm,C^\pm\} \stackrel{I_0}
\longrightarrow  &\{\tilde A^\mp,\tilde B^\mp,C^\pm\}
\\[2.ex]
\{A^\pm,B^\pm,C^\pm\} \stackrel{I_1}
\longrightarrow  &\{\tilde A^\pm, B^\pm,\tilde C^\pm\}
\\[2.ex]
\{A^\pm,B^\pm,C^\pm\} \stackrel{I_2}
\longrightarrow  & \{A^\pm,\tilde B^\pm, \tilde C^\mp\}
\end{array}
\]
where, for instance, the sets $\{A^\pm,A\}$ and $\{\tilde A^\pm,\tilde A\}$ close the two commuting 
 Lie algebras $su(2)$ of section~\ref{su3hamiltoniansystem}. The explicit expression for the new operators (labelled with a tilde) can be easily obtained in the same way as it was done in (\ref{ta01}). The set of all the generators obtained in this process close the Lie algebra of rank 3,  $so(6)$.  In the eigenfunction space it is enough to consider three independent diagonal operators $\{L_0,L_1,L_2\}$ defined by
\[\label{lll}
L_i\, \Psi_{(l_0,l_1,l_2)} = l_i \,\Psi_{(l_0,l_1,l_2)}
\]
The Hamiltonian can be expressed in terms of the  $so(6)$-Casimir operator    by means of the `symmetrization' of the $u(3)$-Hamiltonian (\ref{esp})
\[\label{cass}
\begin{array}{ll}
H_{so(6)}&= \frac18\left(H_{u(3)} + \sum_j I_j\, H_{u(3)}\, I_j
+\sum_{j\neq k} I_jI_k\, H_{u(3)}\, I_jI_k + I_0I_1I_2\, H_{u(3)}\,I_0I_1I_2\right)\\[2.ex]
&=\{A^+,A^-\}+\{B^+,B^-\}+\{C^+,C^-\}\\[2.ex]
& \qquad +\;
\{\tilde A^+, \tilde A^-\}+\{\tilde B^+, \tilde B^-\}+
\{\tilde C^+, \tilde C^-\}+
{L_0}^2+{L_1}^2+{L_2}^2 +\frac{41}{12}
\end{array}
\]
Henceforth we remove the subindex `$so(6)$' of the Hamiltonian.
 
The intertwining generators of $so(6)$ give rise to larger
three-dimensional Hamiltonian hierarchies
\[
\{H_{(l_0+m+p,l_1+m-n-p,l_2+n)} \},\qquad m,n,p\in\Z
\]
each one including a class of the previous ones coming from $u(3)$.
The eigenstates of these Hamiltonian hierarchies can be classified
in terms of $so(6)$-representations. Let us fix our attention in
those determined by the $so(6)$ IUR's. These IUR's are build
from the fundamental states annihilated by the simple roots
$A^-,C^-,\tilde A^-$
 \[\label{fund} 
 A^-\psi_\ell^{(0)} =
C^-\psi_\ell^{(0)} =\tilde A^-\psi_\ell^{(0)} =0 
\] 
The equations
for the operators $A^-$ and $C^-$ have been used in (\ref{ac}),
while the one for $\tilde A^-$ were already applied in (\ref{fuu}).
Therefore, the wavefunctions of the highest weight vectors take the
form 
\[\label{groundd}
 \Psi^{(0)}_\ell(\phi_1,\phi_2)= N
\,\cos^{1/2}\phi_1\;\sin^{1/2}\phi_1\;\cos\phi_2\;\sin^{l_2+1/2}\phi_2
\]
characterised by the eigenvalues of the diagonal operators,
\[\label{dia} 
L_0\, \Psi_{\ell} = L_1\, \Psi_{\ell} = 0,\qquad L_2\,
\Psi_{\ell} = n \,\Psi_{\ell},\qquad n\in\Z^+
 \]
 This fundamental state is invariant under the inversions $I_0$ and $I_1$, and the
representation, so obtained, is also invariant under $I_2$. Thus,
in this way we arrive at two classes of symmetric IUR's of $so(6)$
\begin{itemize}
\item[(a)]
$(l_0=0,\;l_1=0,\;l_2=0)$;  $\{H_{(m+p,m-n-p,n)}\},\   n,m,p\in \Z $
(even IUR's)

\item[(b)]
$(l_0=0,\;l_1=0,\;l_2=1)$;  $\{H_{(m+p,m-n-p,1+n)}\},\   n,m,p\in \Z$
(odd IUR's)
\end{itemize}
Each of these IUR's is described in the parameter space by an octahedral lattice of points such that it will include an iso-energy $su(3)$ (or ${}_isu(3)$) series of representations, quoted in the above subsection, which correspond to parallel exterior faces of the octahedron and some of its sections. Such sections are determined by the values of the diagonal operator $D$ (or $D_i$) whose values fix the corresponding $u(3)$-representations.

\begin{figure}[htp]
\centerline{
\includegraphics[scale=0.75]{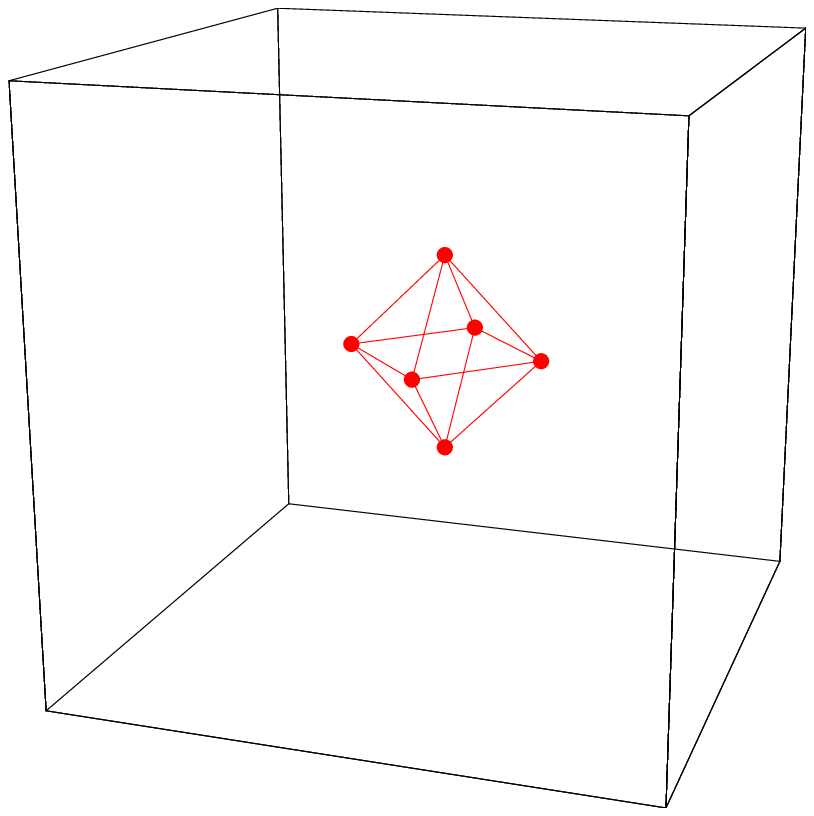}\hskip1.5cm
\includegraphics[scale=0.75]{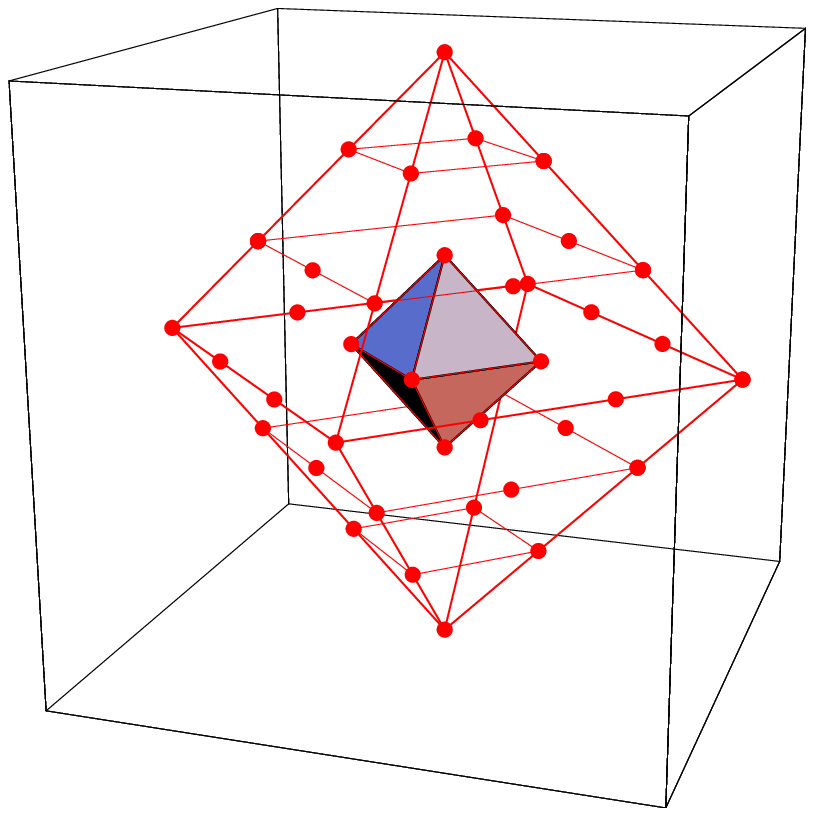}}
\caption{Plot of the points representing the states of two odd IUR's with $q=1$ (left) and $q=3$
(right). The  6 $(q=1)$-eigenstates share the energy
$E=\frac52\cdot\frac32$. The  50 $(q=3)$-eigenstates share the
energy $E=\frac72\cdot\frac52$ (the points corresponding to $q=3$ include those of $q=1$,
of the inner octahedron, which are doubly degenerated.} \label{fig1}
\end{figure}
\begin{figure}[htp]
\centerline{
    \psfrag{l0}{$l_0$}
    \psfrag{l1}{$l_1$}
    \psfrag{l2}{$l_2$}
\includegraphics[scale=0.75]{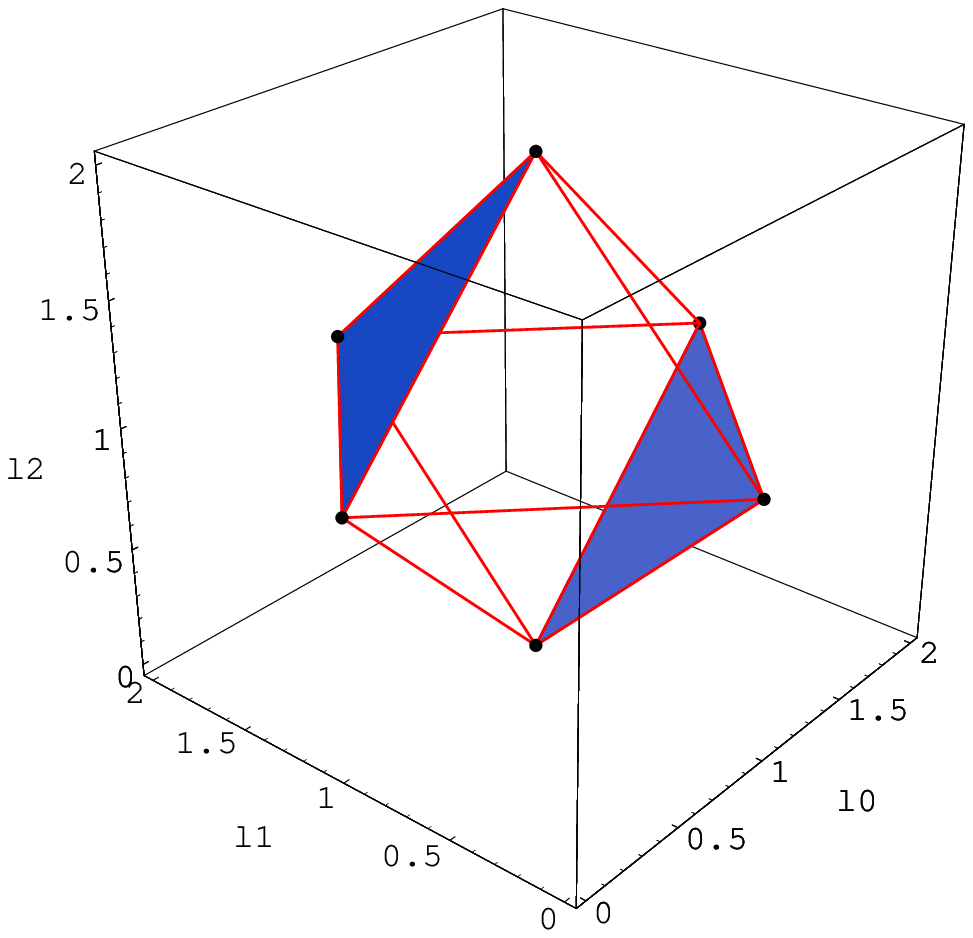}\hskip1.5cm
\includegraphics[scale=0.75]{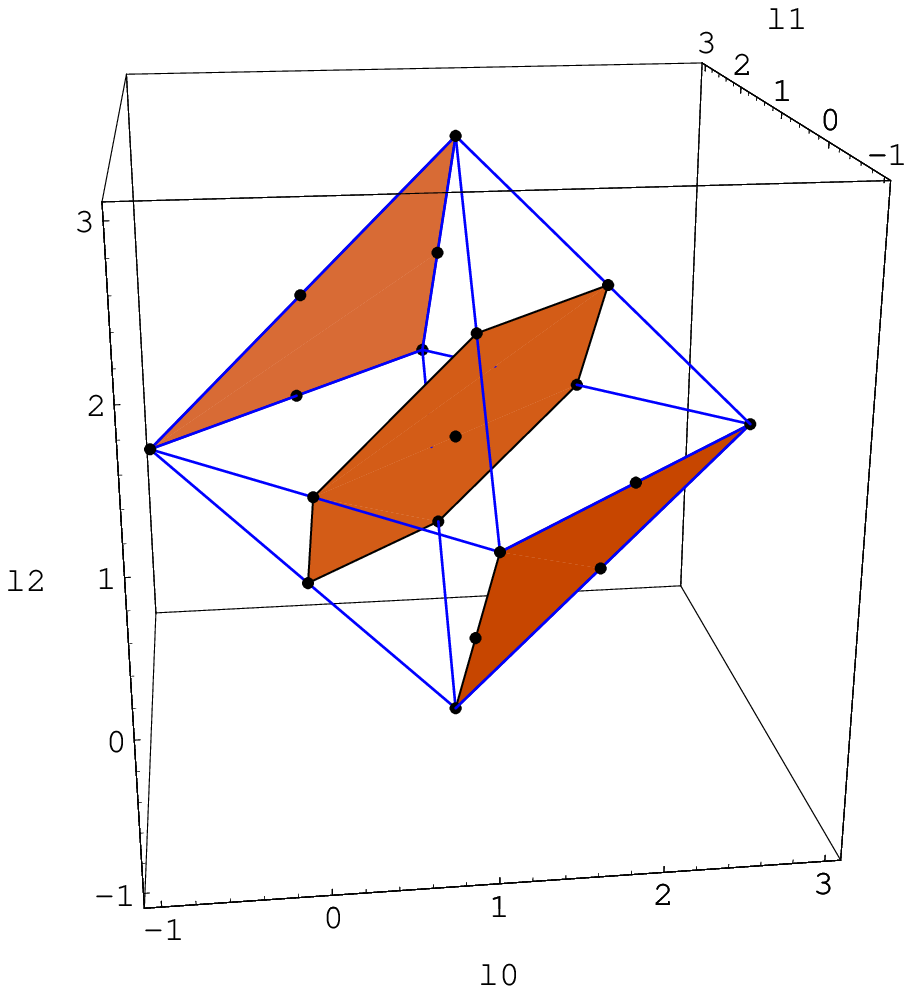}}
\caption{The figure on the left is for the $q=1$ IUR of $so(6)$ where the triangular oposite faces correspond to two IUR's of $su(3)$. The figure on
the right correspond to the the points of a $q=3$ IUR of $so(6)$. The three sections describe three IUR's of $su(3)$.}\label{fig2}
\end{figure}

For instance, the $so(6)$-representation labelled by $n=1$, corresponding to the odd hierarchy, includes the first $su(3)$-series, $(1,0)$ and $(0,1)$ described by the opposite faces of an elemental octahedron.
The $so(6)$-representation of the even hierarchy  fixed by $n=2$ includes the $su(3)$-series made of $(2,0)$, $(1,1)$, and $(0,2)$. Those associated to $(2,0)$ and $(0,2)$  correspond to opposite triangular faces, while $(1,1)$ is described by the parallel hexagonal section through the origin. These features can be better appreciated in Figures~\ref{fig1} and~\ref{fig2}.

In general, the $so(6)$ IUR's  fixed by the parameter $q$ will include the iso-energy series of the  $su(3)$-representations labelled by $(m,n)$ with $m+n=q$. This is the degeneration explained by the larger algebra $so(6)$.
A similar discussion can be done with respect to the representations of the 
$su(2)\oplus \widetilde{su}(2)$ subalgebra. They can be identified with square sections of the octahedron.

\sect{Eigenstates and factorizations} \label{egeinstatesandfactorizations}

Let $H_\ell$ and $H_{\ell'}$ be  two Hamiltonians  related by
means of a differential operator $X$ in the following form
\be\label{twin}
X\, H_{\ell} = H_{\ell'}\, X \Longrightarrow
X^\dagger\, H_{\ell'} = H_{\ell}\, X^\dagger
\ee
where the dagger denotes adjoint differential operators. Then, it is  said that
$X$ is an intertwining operator connecting
$H_{\ell}$ with
$H_{\ell'}$. 

In a formal way, the eigenfunctions of $H_\ell$ are transformed by
$X$ into eigenfunctions of $H_{\ell'}$, but one must be careful about the
behaviour of some properties, such as square-integrability, singularities, or boundary
conditions, which might be altered by $X$. The intertwining problem just as
introduced in (\ref{twin}), which applies to the $u(3)$-system of section~\ref{dynamicalsymmetries},
takes into account shape invariance, in the sense that the partner Hamiltonian
$H_{\ell'}$ differs from the initial
$H_{\ell}$ simply by changing the values of the parameters: $\ell\to \ell'$.  In general, shape invariance leads to an algebraic structure of the intertwining operators as it happens in our present case.

Now, we will discuss in this section the form of the $su(3)$ intertwining operators of 
section~\ref{dynamicalsymmetries} and its relation to certain eigenstates (similar
considerations also apply  to  $so(6)$). First of all, note that we
can write such operators (see expressions (\ref{als}), (\ref{bls}) and
(\ref{cls})) as
\be\label{fields}
A_{\ell}^\pm = a^\pm + \a_{\ell},\qquad
B_{\ell}^\pm = b^\pm + \b_{\ell}, \qquad
C_{\ell}^\pm = c^\pm +  \g_{\ell}
\ee
where $a^\pm, b^\pm, c^\pm$ stand for vector fields (expressed, for instance, in the
variables
$\phi_1,\phi_2$) defined on the sphere and $\a_\ell, \b_\ell, \g_\ell$ design functions also defined on the sphere. Notice that
\[
a^+=-(a^-)^\dagger=J_2,\qquad
b^+=-(b^-)^\dagger=J_1\qquad
c^+=-(c^-)^\dagger=J_0
\]
where $J_0,J_1,J_2$ close the  rotation algebra $so(3)$. Moreover,  taking the hermitian conjugate we have made use of  the invariant measure on the sphere.
If we write the Hamiltonians in the
hierarchy displaying the kinetic (or free) part and the potential as
\[
H_\ell = H^{({\rm kin})} + V_\ell
\]
we see that the vector  fields originate the kinetic term, i.e.
\[
H^{({\rm kin})}=a^+a^- + b^+ b^- + c^+ c^- 
\]
and the components
$\a_\ell, \b_\ell, \g_\ell$ (defined on the sphere)
 give rise to the potential $V_\ell(\phi_1,\phi_2)$, labelled by the
parameters
$\ell\equiv(l_0,l_1,l_2)$. Substituting (\ref{fields}) in
the Hamiltonian (\ref{esp}) and taking into account (\ref{cas}), we
get the expression
\be\label{cua}
V_\ell(\phi_1,\phi_2) = (\a_\ell)^2 + (a^+\a_\ell) + (\b_\ell)^2+
(b^+\b_\ell) + (\g_\ell)^2 + (c^+\g_\ell) +\l_\ell
\ee
where $\l_\ell$ is a number depending on $l_0,l_1,l_2$. Equation (\ref{cua}) can be considered as a nonlinear partial differential
equation linking the unknowns $\{\a_\ell,\b_\ell,\g_\ell\}$ with the
potential, in a quite similar way to the Riccati equation for the superpotential
$\om$ in the one-dimensional Schr\"odinger equation. For this reason, we
sometimes will refer to
$\{\a_\ell,\b_\ell,\g_\ell\}$ as
superpotential functions. This is in agreement with a more general result \cite{kuru1,kuru2} where the first order intertwining is built by `dressing' the symmetries of the Laplacian operator with certain functions.

The basic property of the one-dimensional
superpotential $\om$ was that it could be considered as the logarithmic
derivative of a Hamiltonian eigenstate (see (\ref{sp1})). Here, we have something similar 
with respect to the superpotentials
$\{\a_\ell,\b_\ell,\g_\ell\}$ but first we want to settle this problem
in general terms. If we know an intertwining operator $X$ satisfying
(\ref{twin}) it can help us in computing certain eigenfunctions of $H_\ell$.
Notice that if we define the kernel, $\mathcal{K}_X$, of $X$ as the
linear manifold of wave-functions annihilated by $X$,
\[
X\,\psi =0,\qquad \forall \psi \in {\cal K}_X
\]
then, such a space is invariant under the Hamiltonian operator $H_\ell$. Thus,
we can look for eigenfunctions inside ${\cal K}_X$, in general a much simpler problem. But, in the case of $X$ being a partial
differential operator, its kernel includes certain arbitrary functions, so it is
still an infinite dimensional space. This is in sharp contrast with ordinary
first order differential operators where the kernel is one-dimensional.

Another option we have at hand is the following. The intertwining relation
(\ref{twin}) implies the commutation
\[
X^\dagger X\,  H_\ell = H_\ell\,  X^\dagger X
\]
This means that we can look for eigenfunctions of $H_\ell$ inside any
eigenfunction space of $X^\dagger X$, not necessarily that one annihilated by $X$, as was the case just considered above.
In this case, however, a similar expression to  (\ref{sp1}) in terms of such eigenfunctions is no longer valid for $\om$.
When we know several intertwining
operators, as in the present case, we can apply them in different ways according to the above comments.

\begin{itemize}
\item[{i)}]
{\it Superpotentials associated to a global fundamental eigenstate of $\{A^-,B^-,C^-\}$.}
We consider the intersection of the kernels of all the intertwining operators.
Assuming that this subspace is one-dimensional we have just one
eigenstate (up to a factor)
$\Phi_0$ annihilated by all the lowering operators $\{A^-,B^-,C^-\}$. So that, we
obtain the following expressions quite similar to (\ref{sp1})
\be\label{spf}
\a_{\ell}=-\frac{(a^-\Phi_0)}{\Phi_0},\qquad
\b_{\ell}=-\frac{(b^-\Phi_0)}{\Phi_0},\qquad
\g_{\ell}=-\frac{(c^-\Phi_0)}{\Phi_0}
\ee
This mechanism corresponds to the  IUR's characterized in section~\ref{dynamicalsymmetries}.
\item[{ii)}]
{\it  Superpotentials associated to a partial fundamental eigenstate.}
If the above subspace is the trivial null space, we can still restrict ourselves to the kernel
subspace of anyone of the intertwining operators, for example $A^-$.  Thus, let $\Phi$ be
an eigenfunction of
$H_\ell$ with
$\Phi\in {\cal K}_{A^-}$, i.e.
$A^-\Phi=0$. This allows us to set
$$
\a_{\ell}=-{(a^-\Phi)}/{\Phi}
$$
From this equation we can also separate variables in
$\Phi$. So that,  the eigenfunction equation $H \Phi = E \Phi$ leads to a second
order ordinary differential equation whose solution can be
easily obtained.

However, we must outline that in this case the remaining superpotential functions
$\b_{\ell},\g_{\ell}$ have not a simultaneous expression (\ref{spf}) in terms of
the same $\Phi$, they need different eigenfunctions.
Under this point of view, section~\ref{su3hamiltoniansystem}
 constitutes an illustration of how this option leads to eigenfunctions separated in the variables $\phi_1,\phi_2$.

\item[{iii)}]
{\it Other excited eigenstates.}
The second option is to solve, for instance, the eigenvalue problem $A^+ A^- \Phi = \a\, \Phi$, requiring at the same time $\Phi$ to be also a  Hamiltonian eigenfunction. In terms of the ambient coordinates $s_0,s_1,s_2$ this equation  is (see also \cite{pogosyan})
\[
\left\{-\left( s_1 \frac{\p}{\p s_0}- s_0 \frac{\p}{\p s_1} \right)^2 + (l_0-1/4)\frac{s_0^2+s_1^2}{s_0^2} + (l_1-1/4)\frac{s_0^2+s_1^2}{s_1^2}\right\} \Phi = \a\, \Phi
\]
The same procedure can be applied with other more general sets of operators commuting with
the Hamiltonian. For instance, we can diagonalize $H$ inside the
subspace
\[
\left(e_2\,A^+A^- + e_1\,B^+B^-+ e_0\,C^+C^-\right) \Phi = \a\, \Phi
\]
where the $e_i$'s are constant coefficients.
This leads to eigenfunctions
separated in elliptic  coordinates, that we do not consider here
\cite{pogosyan}.
\end{itemize}

\sect{Conclusions}

We have shown how to deal with the $u(3)$ (and the
general $u(n)$ case \cite{winternitz,COR96} follows the same pattern~\cite{olmo98}) analog of a class of
factorizable one-dimensional potentials with underlying dynamical algebra
$u(2)$.
The higher rank systems in consideration are well known inside the class of
superintegrable Hamiltonians and, of course, our objective was not
to compute original eigenfunctions. Our interest was to apply a different
point of view  to understand some properties in a new context. For instance, the
classification of the irreducible representations of $su(3)$ in series
corresponding to $so(6)$-octahedrons, and the relations involved in this framework
is a non trivial result that could be best appreciated inside the intertwining technique. The relation of the unitary representations with an special
form of the superpotential functions, or the separable eigensolutions determined in terms of intertwining operators clarifies some of the known procedures.

We have seen how the elements of
one-dimensional factorizations must be adapted to the new context. For example, the
relation of superpotentials and a whole class of eigenfunctions (not just one),
the expression of the Hamiltonian operator is not just a simple
factorization, the lattice of states must be drawn in a three-dimensional
space, etc.

There are several problems that can be adressed using the present procedure.
The systems underlying noncompact algebras $u(p,q)$, inhomogeneous Lie
algebras  $\overline{iu}(p,q)$ and contracted algebras \cite{contracciones} are among the first
applications that we expect to report in a near future. But, in general, any
other integrable Hamiltonian system will allow for this treatment, with or
without variable separation. This application would be of most interest.

\section*{Acknowledgments}
This work has been partially supported by DGES of the
Ministerio de Educaci\'on y Ciencia of Spain under Projects BMF2002-02000 and FIS2005-03989
and Junta de Castilla y Le\'on (Spain) (Project VA013C05).




\begin{thebibliography}{20}

\bibitem{infeld} L. Infeld and T.E. Hull,  {\it Rev. Mod. Phys}.  {\bf 23} (1951) 21.

\bibitem{kuru1}
  \c{S}. Kuru, A. Te\v{g}men and A. Ver\c{c}in,
{\it  J. Math. Phys}. {\bf 42} (2001)  3344.

\bibitem{kuru2}
B. Demircio\v{g}lu,  \c{S}. Kuru, M. \"{O}nder and A. Ver\c{c}in,
 {\it  J. Math. Phys}. {\bf 43} (2002)  2133.

 \bibitem{samani}
K.A. Samani and  M. Zarei,
{\it Ann. Phys}. {\bf 316} (2005)  466.

\bibitem{ranada}
 M.F.~Ra\~nada, {\it J. Math. Phys}. {\bf 36} (1995) 3541;
 {\bf 38} (1997) 4165;
 {\bf 40} (1999) 236;
 {\bf 41} (2000) 2121.
 
\bibitem{santander}
 M.F.~Ra\~nada and M.~Santander, {\it  J. Math. Phys}. {\bf 40} (1999)
    5026;   {\bf 43} (2002)  431;  {\bf 44} (2003)  2149.

\bibitem{ioffe}
F. Cannata, M.V. Ioffe and D.N. Nishnianidze, {\it Phys.  A} {\bf 35}  (2002) 1389.

\bibitem{evans} N.W. Evans, {\it Phys. Rev}. {\bf 41} (1990) 5666;
{\it Phys. Lett}. {\bf 147A} (1990) 483;  {\it J. Math. Phys}. {\bf 32} (1991)
3369.

\bibitem{winternitz} M.A. del Olmo, M.A. Rodr\'{\i}guez and P. Winternitz,
{\it J. Math. Phys}. {\bf 34} (1993) 5118.

\bibitem{calzada}
J.A. Calzada, M.A. del Olmo and M.A. Rodriguez 
{\it J. Math. Phys}. {\bf 40} (1999) 88.

\bibitem{pogosyan} E.G. Kalnins, W. Miller and G.S. Pogosyan, {\it J. Math. Phys}. {\bf
37} (1996) 6439.

\bibitem{fernandez}
D.J.~Fern\'andez C., J.~Negro and M.A.~del~Olmo, {\it Ann.  Phys}. 
{\bf 252}  (1996)  386.

\bibitem{refined00}
J. Negro, L.M. Nieto and O. Rosas-Ortiz, {\it J. Phys. A}  {\bf 33}
(2000) 7207.

\bibitem{barut}
A.O. Barut, A. Inomata and R. Wilson, {\it J. Phys. A} {\bf 20} 
(1987) 4075;  {\it J. Phys. A} {\bf 20}  (1987) 4083.

\bibitem{quesne}
A. del Sol Mesa, C. Quesne and Yu F. Smirnov,  {\it J. Phys. A} {\bf 31} 
(1998) 321.

\bibitem{dutt}
M. Dutt, A. Gangopadhyaya, C. Rosinaru and U. Sukhatme,  {\it J. Phys. A}   {\bf 34} 
(2001) 4129.

\bibitem{confluent}
J. Negro, L. M. Nieto and O. Rosas-Ortiz, {\it J. Math. Phys}. 
{\bf 41}  7964
(2000).

\bibitem{COR96} J.A. Calzada, M.A. del Olmo and M.A. Rodr\'{\i}guez,
{\it J. Geom. Phys}. {\bf 23} (1997) 14.

\bibitem{olmo98} J.A. Calzada, J. Negro and M.A. del Olmo,
Quantum superintegrable  Hamiltonian
systems in {\it Proceedings of the 5th Wigner Symposium}, pp. 233. 
World Scientific, Singapore (1998). 


\bibitem{contracciones}
 J.A. Calzada, J. Negro, M.A. del Olmo, M.A. Rodr\'{\i}guez,
 J. Math. Phys. {\bf 41} (1999)  317.


\end{thebibliography}
\end{document}